# Voltage Stabilization of A DC-Microgrid Using ANFIS Controller Considering EVs, DER, and Transient Storage

Hussein Zolfaghari[1], Hossein Karimi[2], Dr. Hamidreza Momeni[3]
Hussein.zolfaghari@modares.ac.ir[1], Hossein.karimi@ucalgary.ca[2], momeni_h@modares.ac.ir[3]

**Abstract:** In this paper, a DC microgrid will be considered to optimize the operation of this microgrid under a combination of Fuzzy and metaheuristic algorithms.

## 1. Introduction

With the increasingly development of technology, a microgrid play a pivotal role in the energy management part, due to the advantages of DC MGs such as reduced losses and easy integration with energy storage resources, DC MGs pave the way of expand usage of such a beneficial plants[1, 2]. Power systems are the collection of energy resources, including loads, generation units, power conversion units, and storage devices [3, 4]. EVs gradually increased since a few years ago as a storage part of MGs and as a generation unit during shortage of energy for responding demands [5], furthermore, the centralized generation model is being gradually replaced by a distributed generation model [6]. In addition, not only do microgrids improve flexibility of the grids but also increase system reliability [7, 8]. Although microgrid provides power system with noticeable features, it brings complexity in power system control and increases costs of electricity balance and support services [8].

Maintaining a storage in DC MGs to supply critical loads when MG faced with shortage of produced energy by RESs is one of the great importance in such an isolated MGs, because the presence of battery of EVs during blackouts is directly related to its bus voltage stabilization as there is not any generation units or storage. The variation in MGs have very fatal effects, voltage variation might trigger protection devices and disconnect DERs within the MG. Centrally controlled MGs (CCMGs) type is dependent RESs, storage, and controllers. Therefore, it is very important to take care of such storage, DERs, and Control units[9, 10].

The invention of new technologies in renewable energy and distribution generation have resulted in lower cost and emission. The introduction of microgrids in power system facilitates the integration of renewable energy into power grid [11]. Due to the stochastic nature of renewable energy, energy storage are necessary to compensate short- and long-term energy variations [12, 13]. A step change in load demand can be considered as a short-term energy variation, whereas changes in produced energy in a long time can be considered as a long-term energy variation [14, 15]. Solar photovoltaic (PV) technology is emerging as a leading source of electricity generation. PV-based generation facilities are vulnerable to faults that, if mismanaged, can impact the supply of energy to the grid and risk system disruption[16]. Convex optimization will improve the results[17]. Shared control method is another method of increase the productivity of stabilizers[18]. Heat transfer will improve the results of optimization in an electrical system[19].

Renewable energy resources are proposed in many papers for demand response. For example, in [20] a Linear programming method are proposed for optimizing the usage of such resources. This paper proposes an important role of electrical vehicles for energy storage and photovoltaic for energy generation. An MLIP cost function is proposed in this paper, thereby optimizing process is easier.

In this research [21] the author states a MMPC solution for the issue in hand in this paper. The paper is about a biological system and implementing a new control method. In this research they improve the results considering the side effect of different control parameters. In DC-MG it will improve the results if the side-effects of elements has been considered.

Solar systems are a type of cost-efficient energy resources in this area. Using such systems has a great number of pros and cons. In [16] the author investigate different bad condition for a solar system. The structure and characteristics of such system will be considered as a sample to show how using solar systems will be secure.

Distributed generators are integrated with storage facilities and loads to form an autonomous DC microgrid. To overcome the control challenges associated with coordination of multiple batteries within one stand-alone microgrid, control layer is established by an adaptive voltage-droop method aimed to regulate a common bus voltage and to sustain the States of Charge (SOCs) of batteries close to each other during moderate replenishment [15]. In [22] incremental conductance algorithm is used to track maximum power from photovoltaic power plant in a DC microgrid. Mathematical models of fuel cells, photovoltaic, and ultracapacitor converters for the control of power plant are described in [23]. In [24], a parametric programming-based approach for the energy management in microgrids is proposed. A parametric mixed-integer linear programming problem is, in addition, formulated for a grid-connected microgrid with photovoltaic, wind, load demand, and energy storage facilities. It is easy enough to conclude that the proposed method is able to model uncertainties effectively, in wind and solar energy resources.

## 2. problem definition

In this paper, a DC MG consisting of a stochastic power source—DERs , a stochastic impedance load, a fixed impedance ballast, and a stabilizer unit is considered—consisted of three branches, namely battery of EV, super capacitor, and over voltage discharge—to protect from EV's battery and super capacitor from overcharge. A central fuzzy inference controller is applied to regulate DC bus voltage, achieving power sharing of batteries and super capacitor, and controlling current stabilizer unit. Fuzzy rules are defined based on researchers' experience and then Particle Swarm Optimization (PSO) is used to optimize fuzzy rules and fine tune fuzzy membership functions. It is shown that optimized fuzzy controller in comparison to the conventional PI controller is more capable to regulate DC voltage while increasing operating life of EV's battery, as a main storage system. Furthermore, fuzzy logic can execute a balancing effect between storage elements and transfer excess energy in one element to another, which having any energy in that of. This feature can easily be introduced, applied, and optimized by fuzzy logic controller while a PI controller, requiring several additional control loops and algorithms for such feature, is not able to do this.

The rest of paper organized as follows: section III presents the DC microgrid case study model. Fuzzy logic inference system and PSO optimization algorithm are introduced in section IV. The results are presented in section V, and section VI concludes the paper.

## 3. DC Microgrid Configuration

The simplified structure of the DC MG with a variable resource, a variable load, a stabilizer. The models of a stochastic power source, a stochastic load, a stabilizer, and a ballast load are illustrated in current section[25].

### 3.1. Stochastic Power Source Model

A maximum power point tracking controller is considered in this study. A pseudo-random number generator provides a target power and a boost converter tracks it to model the stochastic characteristic of the power resource, used in this research. A boost converter duty cycle is defined related to the target power.

### 3.2. Stochastic load model

To model a stochastic load, a pseudo-random number is generated to define power, drawing from the grid. Then, equivalent resistance is calculated and imposed to the grid.

### 3.3. Stabilizer model

Two important sections are considered in stabilizer unit. One section should be considered as power resources to balance the energy, so that of includes battery and ultracapacitor. Also, in the case of excess energy, a dissipating element should be considered to draw the excess power, especially when the battery and ultracapacitor are fully charged. Therefore, stabilizer unit includes a battery, an ultracapacitor, and a dissipating element. Dissipating element is also known as Over Voltage Discharge (OVD).

### 2.4. Ballast load

Since there exist some boost converters in the DC MG, so it is an appropriate choice to intend a minimum load at all times on DC MG. A boost converter with no load can increase voltage significantly and become unstable and damage itself. Therefore, a large-valued resistor is imposed on the grid.

## 3. Control Structure

### 3.1. Conventional PI controller

To control the voltage of the main bus of DC MG, charge, discharge of the battery, charge, discharge of the ultracapacitor, and define the duty cycle of the OVD phase, two cascade PI controllers have been considered for each of phases. In outer control loop, bus voltage error is given to the PI controller and output of the PI controller provides current reference for the battery, the ultracapacitor, and the OVD phases. Another PI controller is used separately to track the current reference by providing the duty cycle of the converter of battery, ultracapacitor, and OVD phases.

### 3.3. Optimization method

The PSO is chosen as an optimization algorithm, since its results are so accurate and does not need complex calculation [31, 32]. Also, based on previous works on fuzzy optimization, the PSO can optimize fuzzy membership functions more accurately and quickly in comparison to other algorithms [33]. Different literature is explained PSO in detail [34, 35], avoiding to repeat here. Flow chart of optimization is shown in Fig 14. In this part after implementing the PSO algorithm the results of [25, 26] has been improved because by reducing the current, spending by battery of EV's. The cost function which is implemented had something to do with battery life-time. Considering the minimum current spending by EV's these results has been showed in this paper.

The flowchart works as follows:

1. PSO parameters are determined
2. Boundaries of the expectation and the standard-deviation of each membership functions are defined.
3. The Fuzzy Inference System (FIS) is initialized.
4. PSO updates its positions and velocities of each population.
5. PSO runs MATLAB/Simulink and provide it with a new FIS. After simulation, PSO calculates the objective function value.
6. If this new FIS results in a better answer, it is considered as a best FIS up to now.
7. If stop condition is not met, go to step 4.
8. Print the results.

## 4. Fuzzy Training and Numerical Study

As mentioned in the previous sections the DC MG in issue is made of four parts, namely stochastic power resource, stochastic load, stabilizer unit, and ballast resistor[25, 26]. The system is simulated in MATLAB/Simulink. Battery bank are made by connecting four 12 V, 10 Ah unit in series form[36]. Battery voltage changes from 47.2 V to 50.8 V. The voltage “47.2 V” is considered as exhausted resource (0% of SOC) and “50.8” is considered as full charge (100% SOC). The capacity of ultracapacitor is considered as 150F and 54 V, presented in the following two sections. In the first, the initial fuzzy system is optimized based on objective function. In the next section, system is simulated, and results are compared and discussed

### 4.1. Cost function considering EVs' battery lifetime

The optimization objective function consists of two terms. The first term is DC bus voltage error and the second term in absolute value of the battery current of EV. The first term is essential, since the main goal of fuzzy controller is control of DC bus Voltage. The second term is also important because ultracapacitor is less sensitive to charge and discharge stress in transient time in comparison to the EV's battery, and it has been tried to impose this transient stress to the ultracapacitor. A piece of information that should bear in mind is that, in this study, the membership functions of the outputs have been tuned, since they play more important role in DC microgrid control. Also, the membership functions of inputs do not need essential modifications, because the real condition fuzzification process have been tried to map here. The training objective function is as follows: This objective function will minimize the voltage ripple and the number of charge and discharge of batter as well. The penalty will define a new constraint that help the system to show a higher level of control for battery lifetime.

The PSO iteration is chosen 100 and population is chosen 60. The simulation time is considered 150 second, in MATLAB scale, producing a change in power 10 second and load changes every 3 seconds.

## V. Conclusion

This paper represents a new control methodology for DC Microgrid control. The inputs of proposed fuzzy controller get four variables, that is the error of bus voltage, the integrated error of bus voltage, the SOC of the battery, and the SOC of the ultracapacitor to define currents of stabilizer units. The simulation has shown the proposed controller is successful in bus voltage regulation. The main contribution of the proposed method in comparison to the others is lower stress on the battery and also proper energy transmission between different storage when one of them is almost full charged and another is completely depleted. Also, the initial fuzzy controller has been tuned by PSO to even improve the results.